\documentclass[prc,aps,superscriptaddress,twocolumn,showpacs,nofootinbib]{revtex4-1}
\usepackage{amsmath}
\usepackage{amssymb}
\usepackage{amsthm}
\usepackage{amsfonts}
\usepackage{graphicx}
\usepackage[usenames,dvipsnames]{color}
\usepackage{afterpage}
\usepackage{mathrsfs}
\usepackage{graphicx,latexsym,epsfig}
\usepackage{mathrsfs}
\usepackage{multirow}
\usepackage{color}
\usepackage{bm} 

\newcommand{\beq}{\vspace{0.5em}\begin{equation}}
\newcommand{\eeq}{\end{equation}\vspace{0.5em}}
\newcommand{\beqn}{\vspace{0.5em}\begin{eqnarray}}
\newcommand{\eeqn}{\end{eqnarray}\par\vspace{0.5em}\noindent}

\newcommand{\bsub}{\begin{subequations}}
\newcommand{\esub}{\end{subequations}}
\newcommand{\br}{\mathbf{r}}

\begin{document}
 \preprint{preprint}

 \title{Quantum and Coulomb repulsion effects on the bubble structures in $^{204, 206}$Hg}

  \author{X. Y. Wu}
  \affiliation{College of Physics and Communication Electronics, Jiangxi Normal University, Nanchang 330022, China}
  \author{J. Xiang}
  \affiliation{School of Physics and Electronic, Qiannan Normal University for Nationalities, Duyun, 558000, China}
 \date{\today}

 \begin{abstract}
 The decreasing proton and charge densities from around 5.0 fm towards the center of $^{204, 206}$Hg are investigated by a covariant density functional theory at the beyond mean-field level.
 The charge-density difference between $^{208}$Pb and $^{204}$Hg is improved significantly and a central depression is still visible in the ground-state density of $^{204, 206}$Hg when the dynamic correlations associated with symmetry restoration and shape mixing are taken into account. For the $0_2^+$ and $2_1^+$ excited states of $^{204, 206}$Hg, their densities remain decreasing from  5.0 fm to around 2.0 fm, but become flat in the interior region. The results show that the bubble structure in $^{204, 206}$Hg within 2.0 fm is mainly attributed to the quantum effect, while that beyond 2.0 fm is formed by the Coulomb repulsion.

 \end{abstract}

\maketitle

 \section{Introduction}
 \label{Introduction}
 Because of the saturation properties of nuclear matter, nuclear density generally takes the form of a Fermi distribution. However, in some light or medium-heavy nuclear systems, the density deviates from this simple behavior because of quantum effects related to the filling of single-particle states with wave functions that have a specific spatial behavior. In this context, $s_{1/2}$ orbits in spherical nuclei have a very peculiar signature, as they are the only ones that contribute to the density at the nuclear center. Depending on whether they are filled or empty, $s_{1/2}$ orbits can generate a central bump in the density as it has been observed for $^{36}$S~\cite{Rychel83}, $^{40}$Ca~\cite{Sick79} and $^{208}$Pb~\cite{Euteneuer78,Cavedon82}, or a central depression in the proton density of $^{34}$Si~\cite{Mutschler16} and $^{204}$Hg~\cite{Burghardt89}. In heavy or superheavy nuclei with a large charge number, the density is prone to take the form of ``wine-bottle" shape as it lowers the Coulomb repulsion. Therefore, the depletion of proton and charge densities in the center, referred as ``bubble", is generally governed by both the quantum effect and the compromise between the large repulsive Coulomb interaction and the attractive nucleon nucleon strong force.

 Since the pioneering work of Wilson~\cite{Wilson46}, the bubble structure in atomic nuclei has attracted much attention~\cite{Campi73,Davis73,Bender99,Decharge99,Decharge03,Afanasjev05,Pei05,Khan08,Grasso09,Wang11,Wang15,Li16,Saxena18}.
 Mean-field methods are the tools of choice for modeling the nuclear density distribution and thus most of the theoretical studies on bubble structure were carried out within this framework. In recent years, the multi-reference energy density functional (MR-EDF) calculations~\cite{Yao12,Yao13} along with {\em ab initio} calculations~\cite{Duguet17} have been carried out for $^{34}$Si and $^{46}$Ar~\cite{Wu14}. It has been found that the beyond mean-field dynamic correlation effects quench or even wash out the depletion at the center of the bubble candidate nucleus. It can be understood that the deformation and shape fluctuation distort the spherical shell structure and bring the $s_{1/2}$ orbits partially filled. Nevertheless, the bubble structure in the ground-state of $^{34}$Si is rather robust in both the MR-EDF and {\em ab initio} calculations and it has been indirectly confirmed from the measured small proton occupancy 0.17(3) of the $2s_{1/2}$ orbit~\cite{Mutschler16}.

 The mercury isotopes around neutron number $N=126$ are good candidates with a bubble structure in medium-mass region based on the following two considerations. On one hand, the proton $3s_{1/2}$ orbit is expected to be filled completely in $^{208}$Pb, and it is expected to be depopulated entirely in $^{206}$Hg. On the other hand, the $N=126$ shell gap is robust to hinder the coupling of ground state to large amplitude collective excitations. Besides, the mechanisms of both quantum effects and Coulomb repulsion are expected to play roles in the formation of bubble structure in the nuclei of this mass region. Therefore, several efforts have been made to devoted into the research on the density distribution of  $^{206}$Hg. The relativistic mean-field (RMF) approach predicted a visible proton hollow in  $^{206}$Hg~\cite{Todd-Rutel04}, which is, however,  not supported by the recent studies with spherical Hartree-Fock-Bogoliubov models~\cite{Wang15,Li16}. It has been pointed out that the small shell gap between 2$d_{3/2}$-$3s_{1/2}$ and the strong pairing correlation annihilates the bubble structure in $^{206}$Hg. In this paper, we are going to revisit this topic with the MR-EDF approach based on a relativistic point-coupling energy functional. A special emphasis will be placed on the changes in the density distribution for the low-lying states of $^{204, 206}$Hg under the perturbation of the dynamic correlations.

 The paper is organized as follows. In Sec.~\ref{Method}, we present a brief introduction of the method. The results on the density distribution in $^{204, 206}$Hg and the discussion on the dynamic correlation effects are given in Sec.~\ref{Results}. The conclusions are drawn in Sec.~\ref{Summary}.

 \section{Method}
 \label{Method}
 The MR-EDF approach that we are using in this work has been introduced in Refs.~\cite{Yao10,Yao13,Yao15}. Here we just give an outline of it. In this approach, the wave functions of nuclear low-lying states  are constructed as a superposition of a set of quantum-number projected mean-field states
 \begin{equation}
 \label{GCMWF}
 \vert \Psi^{JNZ}_\alpha\rangle
 =\sum_\beta f^{J}_\alpha(\beta)\hat P^J_{M0} \hat P^N\hat P^Z\vert \Phi(\beta)\rangle,
 \end{equation}
 where $\hat P^J_{M0}$, $\hat P^N$ and $\hat P^Z$ are the projection operators onto angular momentum, neutron and proton numbers, respectively. $\vert \Phi(\beta)\rangle$s are axially deformed states from the RMF+BCS calculations with a constraint on the mass quadrupole moment $\langle Q_{20} \rangle = \sqrt{\dfrac{5}{16\pi}}\langle \Phi(\beta)\vert 2 z^2 - x^2 - y^2\vert  \Phi(\beta) \rangle$, where the deformation parameter $\beta$ is related to the quadrupole moment by $\beta= \dfrac{4\pi}{3AR^2} \langle Q_{20} \rangle$, $R=1.2A^{1/3}$ with mass number $A$.

 The weight function $f^{J}_\alpha(\beta)$ of the states in Eq.~(\ref{GCMWF}) is determined by the variational principle which leads to the Hill-Wheeler-Griffin equation,
  \beq
  \label{HWE}
  \sum_{\beta'}
  \left[{\cal H}^J(\beta,\beta') -E^{J}_{\alpha}
 {\cal N}^J(\beta,\beta')\right]
   f^{J}_{\alpha}(\beta')=0\, ,
  \eeq
 where the norm kernel ${\cal N}^J(\beta,\beta')$ and the Hamiltonian kernel
 ${\cal H}^J(\beta,\beta')$ are defined as
 \bsub
 \beqn
 {\cal N}^J(\beta,\beta')&=&
  \langle \Phi(\beta) \vert
 \hat{P}^J_{00}\hat{P}^N \hat{P}^Z \vert\Phi
 (\beta')\rangle
 \, , \\
  {\cal H}^J(\beta,\beta')&=&
  \langle \Phi(\beta) \vert \hat H
 \hat{P}^J_{00}\hat{P}^N \hat{P}^Z \vert\Phi
 (\beta')\rangle\, .
 \eeqn
 \esub

 With the wave functions of nuclear low-lying states, one can derive the corresponding density distribution in coordinate space~\cite{Yao15},
 \beqn
 \label{eq_GCMrho}
 \rho^{J\alpha}(\br)
 &\equiv& \langle \Psi^{JNZ}_\alpha\vert \hat \rho \vert \Psi^{JNZ}_\alpha\rangle\nonumber\\
 & = &   \sum_{\beta\beta'}  f^{J}_{\alpha} (\beta')f^{J}_\alpha(\beta)
 \sum_{\lambda}  (-1)^{2\lambda}Y_{\lambda0}(\hat \br) \nonumber\\
 &&\times
 \langle  J0,\lambda 0\vert J 0\rangle  \sum_{K}(-1)^{K}\langle JK,\lambda -K\vert J0\rangle \nonumber \\
        &   & \times
 \int d\hat\br^\prime \rho^{JK0}_{\beta'\beta}(\br^\prime)Y^\ast_{\lambda K}(\hat \br^\prime)\, ,
 \eeqn
 where the $\rho^{JK0}_{\beta'\beta}(\br)$ is defined as
 \beqn
  \label{rho_JKNZ}
  \rho^{JK0}_{\beta'\beta}(\br)
  &=& \dfrac{2J+1}{2} \int^\pi_0 d\theta\sin(\theta) d^{J\ast}_{K0}(\theta)\langle \Phi(\beta^\prime)\vert \nonumber\\
  &&\times
  \sum_{i}\delta(\br-\br_i)e^{i\theta\hat J_y} \hat P^{N}\hat P^{Z}  \vert \Phi(\beta)\rangle \, .
 \eeqn

 The index $i$ in the summation runs over all the occupied single-particle states for neutrons or protons. $\hat\br\equiv(r, \hat \br)$ is the position at which the density is to be calculated and $\br_i$ is the position of the $i$-th nucleon.

 The density in Eq.~(\ref{eq_GCMrho}) contains the information of many deformed mean-field states generated by the collective coordinate $\beta$ and it corresponds to the density in the laboratory frame. The density for the $0^+_1$ ground state can be simplified as
 \begin{eqnarray}
 \label{rho_01}
 \rho^{g.s.}(\br) =  \sum_{\beta'\beta}  f^{0}_1 (\beta')f^{0}_1(\beta) \int d\hat\br\rho^{000}_{\beta'\beta}(r, \hat\br) \, ,
 \end{eqnarray}
 where $\hat \br$ denotes the angular part of coordinate $\br$.

 The charge density is calculated by a convolution of the corresponding proton density with a Gaussian form factor,
 \beqn
 \rho_{\rm ch}(r)
 &=&\frac{1}{a\sqrt{\pi}}\int dr^\prime r^\prime\rho_{p}(r^\prime)\nonumber\\
 &&\times\left[\frac{e^{-(r-r^\prime)^2/a^2}}{r}
 -\frac{e^{-(r+r^\prime)^2/a^2}}{r}\right]\, ,
 \label{rhocha}
 \eeqn
 where the parameter with a proton size $a=\sqrt{2/3}\langle r^2_p\rangle^{1/2}=0.65$ fm is adopted in calculations~\cite{Negele70}.


 \section{Results and discussion}
 \label{Results}

 In the mean-field calculations, the symmetries of parity, $x$-simplex, and time-reversal invariance are assumed. The Dirac equation for single-particle wave functions in each reference state $\vert\Phi(\beta)\rangle$ is solved in a set of three-dimensional harmonic oscillator basis within 14 major shells. Pairing correlations between nucleons are treated with the BCS approximation using a density-independent $\delta$ force implemented with a smooth cutoff factor~\cite{Krieger90}. More details on the techniques adopted to solve the RMF equations have been introduced, for instance, in Ref.~\cite{Gambhir90} and review papers~\cite{Ring96,Vretenar05,Meng06}. In the calculations of kernels, the number of mesh points in the interval $[0, \pi]$ for the Euler angle $\theta$ and gauge angle $\varphi_{\tau}$ are chosen as 14 and 9 in the angular-momentum and particle-number projections, respectively. The Pfaffian method~\cite{Robledo09} is carried out to evaluate the phase of the norm overlap in the kernels.


 \subsection{Bubble structure in ground states}
 \begin{figure}[htbp]
 \centering
 \includegraphics[width=8.5cm]{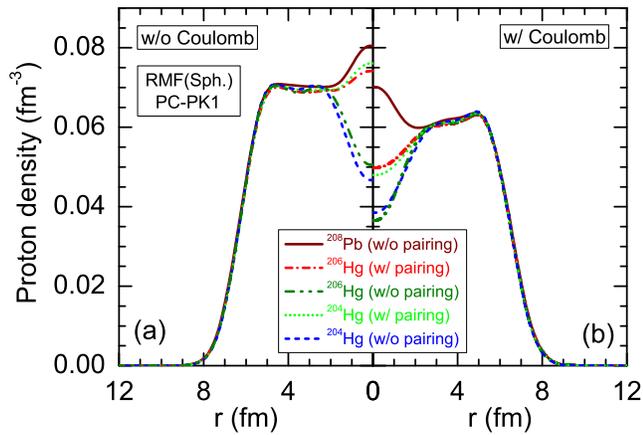}
 \caption{(Color online)
 Comparison between the radial distribution of proton densities (a) without (w/o) and (b) with (w/) Coulomb potential, calculated from the RMF calculations for the spherical states of $^{208}$Pb, $^{206}$Hg, and $^{204}$Hg using the PC-PK1 force.  The results without (w/o) pairing are also given for $^{206}$Hg and $^{204}$Hg. Pairing collapse takes place in the spherical state of $^{208}$Pb labeled by ``w/o pairing''.
 }
 \label{Prodens204206HgPb208}
 \end{figure}

 \begin{figure}[htbp]
 \centering
 \includegraphics[width=16.0cm]{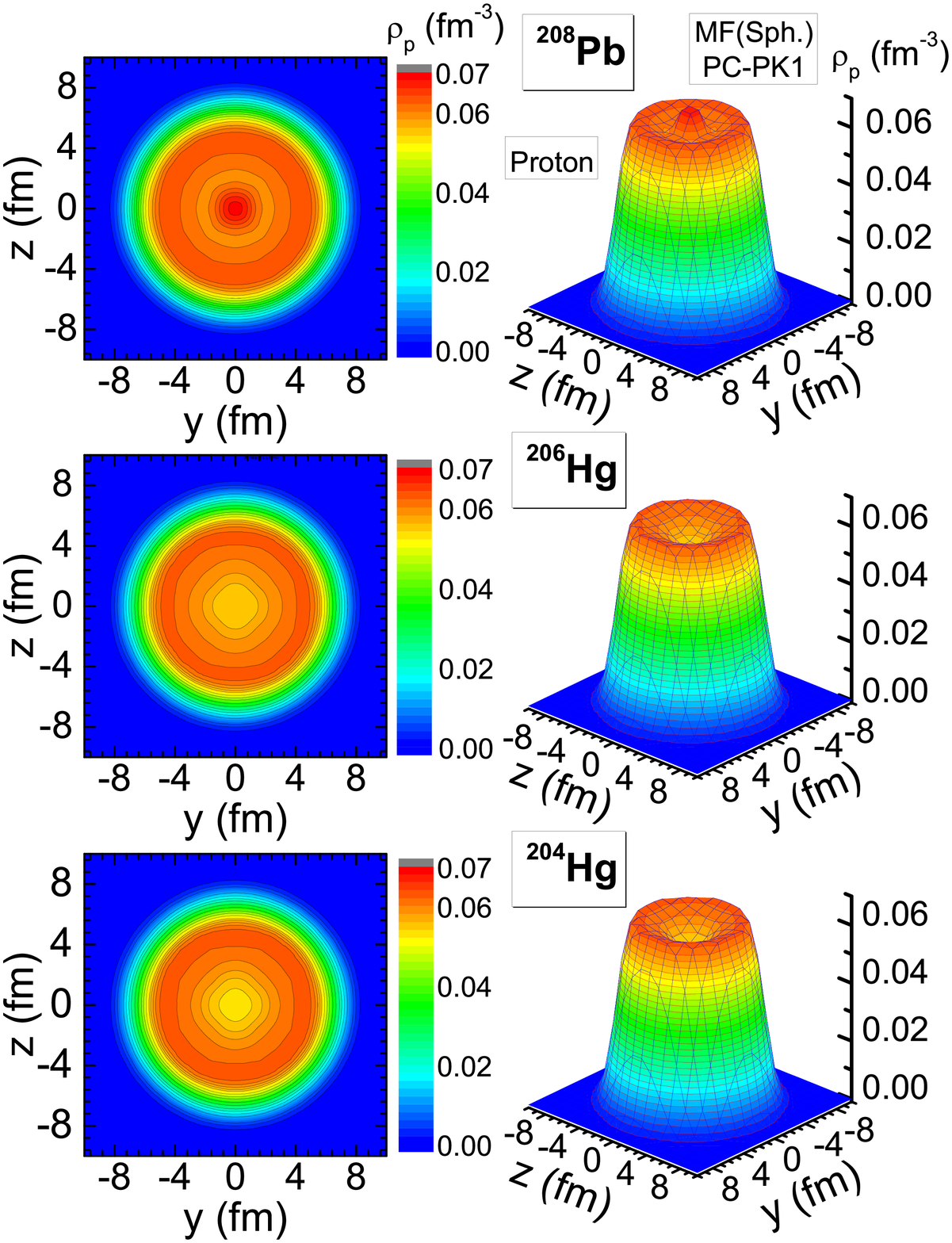}
  \vspace{-11.5cm}
 \caption{(Color online) The proton density distributions (in units of fm$^{-3}$) of the mean-field spherical state in $y$-$z$ plane at $x=0$ fm by the PC-PK1 force for $^{208}$Pb (top), $^{206}$Hg (middle), and $^{204}$Hg (bottom), respectively.}
 \label{23DProdens204206HgPb208}
 \end{figure}

 Figure~\ref{Prodens204206HgPb208} displays the density distributions of protons for the mean-field spherical states of $^{208}$Pb, $^{206}$Hg, and $^{204}$Hg with the PC-PK1 force~\cite{Zhao10}. To examine the effect of Coulomb repulsion on the proton densities, the results from the calculations with or without  the Coulomb potential  are shown  for a comparison. In the realistic calculations with the Coulomb potential,  the proton densities are obviously depressed in the interior region of $^{204, 206}$Hg. The pairing correlations quench the bubble structure significantly by scattering protons onto $3s_{1/2}$ orbit. The  occupation probability of the proton $3s_{1/2}$ orbit in $^{204}$Hg and $^{206}$Hg  is $\sim$ 48\% and $\sim$ 52\% of that in $^{208}$Pb, respectively.  Nevertheless, the central densities are still much depressed. It is shown in Fig.~\ref{Prodens204206HgPb208}(a) that the central depression disappears in the densities from the calculations without the Coulomb potential. It indicates that the central depression shown in Fig.~\ref{Prodens204206HgPb208}(b) is largely attributed to the Coulomb repulsion. The central bump (and central depression) in the proton densities of $^{208}$Pb (and  $^{204, 206}$Hg) can be seen more clearly in Fig.~\ref{23DProdens204206HgPb208}. We also carried out a relativistic Hartree-Bogoliubov calculation with  the PC-PK1 force plus  a separable pairing interaction for $^{206}$Hg. A central hollow is shown as well, even though the central proton value is larger than the value of the RMF+BCS calculations by about 0.01 fm$^{-3}$.

 \begin{figure}[htbp]
 \centering
 \includegraphics[width=8.8cm]{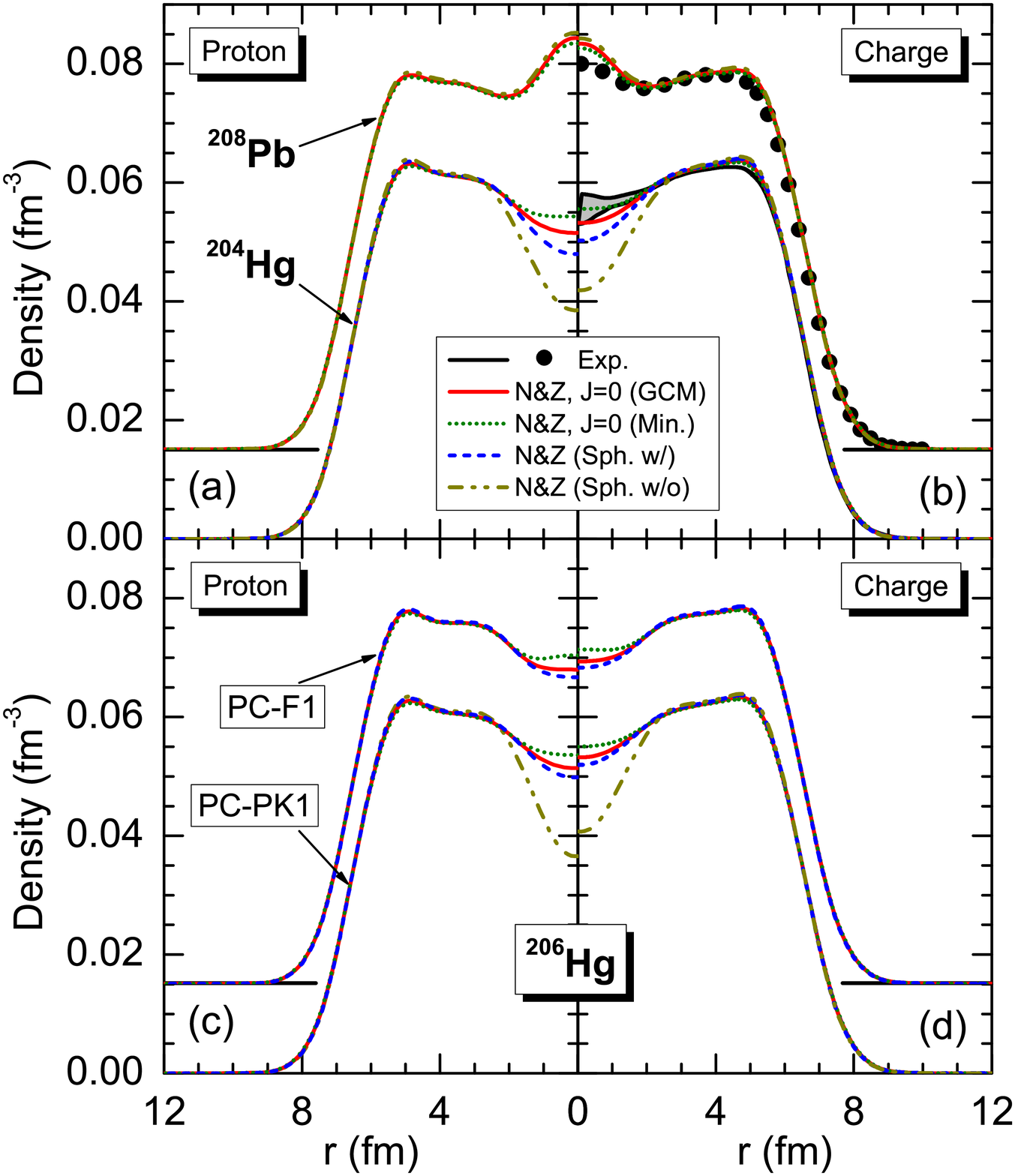}
 \caption{The (a) proton and (b) charge density distributions by the PC-PK1 force for $^{208}$Pb and $^{204}$Hg, as well as the comparison between (c) proton and (d) charge densities for $^{206}$Hg by both the PC-F1 and PC-PK1.
 The labels ``$N\&Z$(Sph. w/o)'', ``$N\&Z$(Sph. w/)'', ``$N\&Z, J=0$(Min.)'', and ``$N\&Z, J=0$(GCM)''
 represent the results based on four different configurations.
 The densities for $^{208}$Pb in (a) and (b), and the density by the PC-F1 for $^{206}$Hg in (c) and (d) have been shifted up by 0.015 fm$^{-3}$. The shadow area  in (b) denotes the experimental uncertainty.
 The experimental data are taken from Refs.~\cite{Euteneuer78,Burghardt89}. See text for more details.}
 \label{dens204HgPb208}
 \end{figure}

 Figure~\ref{dens204HgPb208} shows the proton and charge densities from the calculations based on four different configurations, including the spherical state with or without pairing correlations, the state of the energy minimum on the angular momentum projected energy surface with $J=0$, and the GCM ground state with $J^\pi=0^+$.
 The radial distribution of the charge density for $^{208}$Pb beyond $2.0$ fm is reproduced with  the particle-number conserved spherical state. As the proton $3s_{1/2}$ orbit becomes fully occupied, the central bump is overestimated compared to the data, which is similar to the results from the Skyrme Hartree-Fock calculations~\cite{Bennour89}. The interior charge distribution of $^{208}$Pb is not evidently changed by the effects of static and dynamic quadrupole deformations. For $^{204}$Hg, one can see that the charge density of the particle-number conserved spherical state is much lower than the  data~\cite{Burghardt89}, similar to the result from the relativistic Hartree-Bogoliubov calculations using DD-ME2 force~\cite{Li16}. After taking into account the dynamic correlations in this calculations, the charge density of $^{204}$Hg is reproduced. For $^{206}$Hg, both forces predict almost the same density profiles and the results are similar to that of $^{204}$Hg. It is remarkable that the proton and charge densities gradually decrease from around 5.0 fm towards the center of $^{204, 206}$Hg for the ground states. We note that the wave function of GCM ground state is spread over the range of deformation $-0.3 \leqslant\beta\leqslant 0.3$ with the mean quadrupole deformation parameter $\bar\beta_{01}=\sum_{\beta} \vert g^{J=0}_{\alpha=1}(\beta)\vert^2 \beta\simeq0.02$ for $^{208}$Pb and $^{204, 206}$Hg.

 \begin{figure}[tp]
 \centering
 \includegraphics[width=8.5cm]{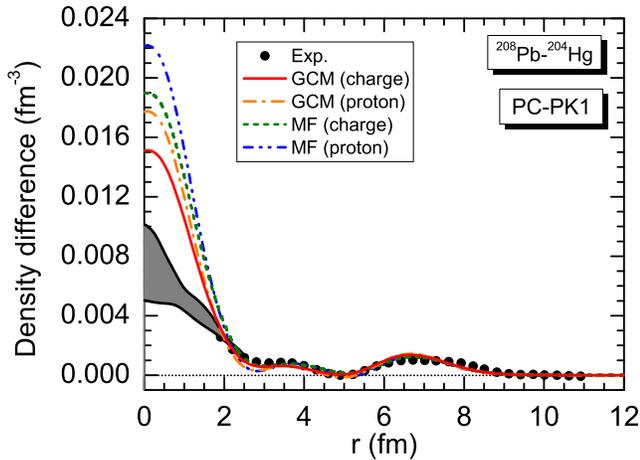}
 \caption{(Color online) The radial distribution of the difference $\Delta\rho(r)$ in the densities between $^{208}$Pb and $^{204}$Hg. The experimental data are taken from Refs.~\cite{Euteneuer78, Burghardt89,Richter03}. The shadow area denotes the experimental uncertainty.}
 \label{dens208Pb-204Hg}
 \end{figure}

 Figure~\ref{dens208Pb-204Hg} displays the difference $\Delta\rho(r)$ in the  proton and charge densities between  $^{208}$Pb and $^{204}$Hg, which reflects mainly the radial distribution of the two protons in $3s_{1/2}$ orbit and has been determined from the measurement on the cross sections~\cite{Cavedon82}.  The main feature of $\Delta\rho(r)$ is reproduced in both mean-field and beyond mean-field calculations. However, the mean-field calculation overestimates significantly the peak value at the center, which is consistent with the results obtained from the Hartree-Fock calculations using finite range effective nucleon-nucleon interactions~\cite{Sick79,Cavedon82,Frois77}.
 After taking into account the effect of dynamic correlations, the central bump is decreased evidently, but not sufficient to reproduce the data.

 \begin{figure}[htbp]
 \centering
 \includegraphics[width=8.5cm]{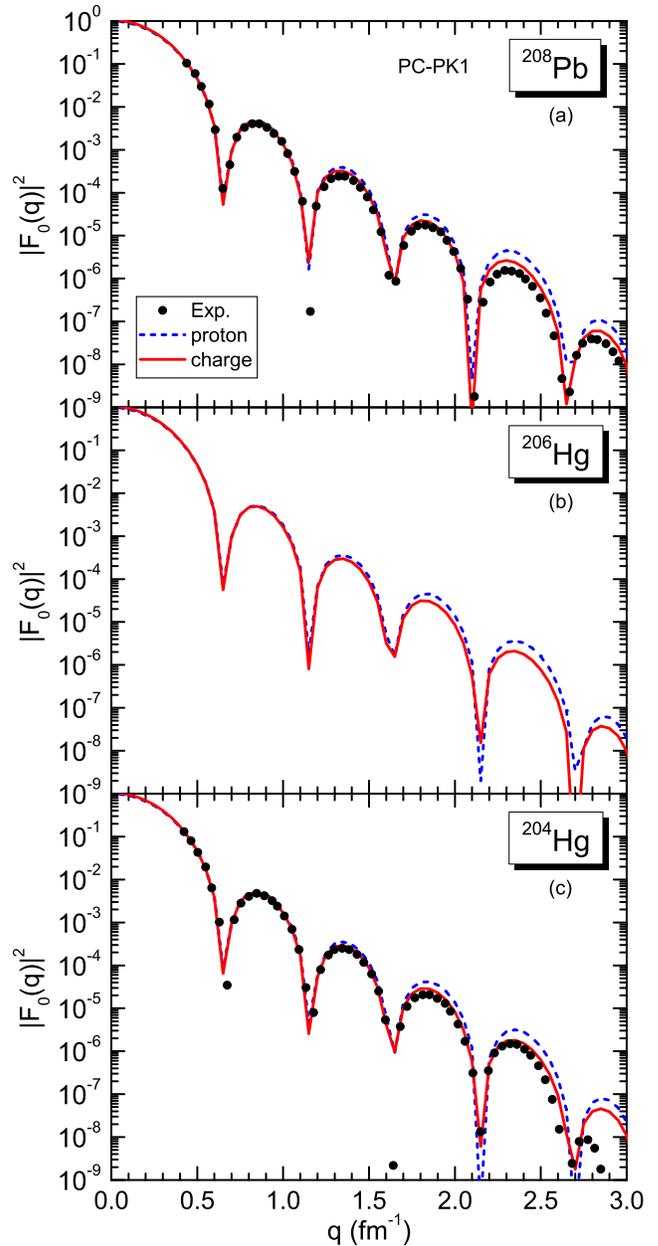}
 \caption{(Color online) Longitudinal  form factor $|F_0(q)|^2$ as a function of the momentum transfer $q$ (fm$^{-1}$) for the electron elastic  scattering of the ground state for (a) $^{208}$Pb, (b) $^{206}$Hg, and (c) $^{204}$Hg, respectively. The experimental data are taken from Ref.~\cite{Richter03}.}
 \label{208Pbfactor}
 \end{figure}

 Figure~\ref{208Pbfactor} displays the  longitudinal Coulomb form factor $|F_0(q)|^2$ corresponding to the electron elastic scattering from the GCM ground state for $^{208}$Pb, $^{206}$Hg, and $^{204}$Hg as a function of momentum transfer $q$, respectively. The form factor $F_J(q)$ is defined as
 \beqn
 \label{formfactor}
 F_J(q) =\frac{\sqrt{4\pi}}{Z}\int_0^\infty dr\,r^2\,\rho^{J\alpha}_{01,J}(r) j_J (qr)\, ,
 \eeqn
 where $j_J (qr)$ is the spherical Bessel function. $\rho^{J\alpha}_{01,J}(r)$ is the reduced transition density~\cite{Yao15}
 \beqn
 \rho^{J\alpha}_{01,J}(r)
 &=&\sqrt{(2J+1)} \sum_{\beta'\beta} f^{J\ast}_{\alpha}(\beta')  f^{0}_{1}(\beta)\nonumber\\
 &&\times \int d\hat{\br}  \rho^{J00}_{\beta'\beta}(\br) Y_{J0}(\hat{\br}).
 \eeqn

 It is shown in Fig.~\ref{208Pbfactor} that  the form factors for $^{208}$Pb and $^{204}$Hg are reproduced rather well when the finite-size effect of protons is taken into account.

 \begin{table}[t]
 \caption{
 \label{tab1} The central and maximal densities (fm$^{-3}$) and the corresponding depletion factors $F_{max}^{\tau}$ and $F_{sat}^{\tau}$ [cf. Eq.~(\ref{Fmax})] for the proton density in $^{206}$Hg from the calculations using the PC-PK1 force. See text for more details.}
 \begin{ruledtabular}
 \begin{tabular}{lllll}
 States                &  $\rho_{cent}^p$ &  $\rho_{max}^p$  & $F_{max}^p$  &  $F_{sat}^p$  \\
 \hline
 $N\&Z$ (Sph. w/o)     &  $0.037 $       &  $0.063 $         &  $0.42 $     & $0.41 $        \\
 $N\&Z$ (Sph. w/)    &  $0.050 $       &  $0.063 $         &  $0.21 $     & $0.20 $        \\
 $N\&Z, J=0$ (Min.)    &  $0.054 $       &  $0.062 $         &  $0.14 $     & $0.14 $        \\
 $N\&Z, J=0$ (GCM)     &  $0.051 $       &  $0.063 $         &  $0.18 $     & $0.17 $        \\
 \end{tabular}
 \end{ruledtabular}
 \end{table}

 \begin{table}[b]
 \caption{
 \label{tab2}
 Same as Table~\ref{tab1}, but for the charge density in $^{206}$Hg. }
 \begin{ruledtabular}
 \begin{tabular}{lllll}
 States                &  $\rho_{cent}^{ch}$ &  $\rho_{max}^{ch}$ & $F_{max}^{ch}$  &  $F_{sat}^{ch}$  \\
 \hline
 $N\&Z$ (Sph. w/o)     & $0.041 $         &    $0.064 $     &  $0.36 $     & $0.34 $       \\
 $N\&Z$ (Sph. w/)    & $0.052 $         &    $0.064 $     &  $0.18 $     & $0.16 $       \\
 $N\&Z, J=0$ (Min.)    & $0.055 $         &    $0.063 $     &  $0.13 $     & $0.11 $       \\
 $N\&Z, J=0$ (GCM)     & $0.053 $         &    $0.063 $     &  $0.16 $     & $0.14 $       \\
 \end{tabular}
 \end{ruledtabular}
 \end{table}

 The depletion factors $F_{max}^{\tau}$ and $F_{sat}^{\tau}$ are often introduced to quantify the bubble structure in the proton and charge density distributions~\cite{Grasso09,Yao12}
 \beq
 \label{Fmax}
 F_{max}^{\tau}\equiv\frac{\rho_{max}^{\tau}-\rho_{cent}^{\tau}}{\rho_{max}^{\tau}}\, , \quad  F_{sat}^{\tau}\equiv\frac{\rho_{sat}^{\tau}-\rho_{cent}^{\tau}}{\rho_{sat}^{\tau}}\, ,
 \eeq
 where $\tau\equiv p$ and $ch$ correspond to the proton and charge, respectively. The values of the central and maximal densities and the corresponding depletion factors for proton and charge are summarized in Tables~\ref{tab1} and~\ref{tab2}.  The saturation density $\rho_{sat}^{\tau}$ is calculated as $\rho_{sat}^{\tau}=(80/206)\times0.16$ fm$^{-3}$ = 0.062 fm$^{-3}$ for $^{206}$Hg. This value is very close to the maximal density.
 As a result, the alternative depletion factors $F_{sat}^{\tau}$ are approximately equal to the values of $F_{max}^{\tau}$. This result is different from $^{34}$Si in which the values of $F_{sat}$ are much smaller than those of $F_{max}$~\cite{Yao12}. Moreover, we note that the correlations quench the central depression, but do not change the maximal densities. In short, the results show that the bubble structure is still survival with the presence of both static and dynamic deformation effects.

 \begin{figure}[htbp]
 \centering
 \includegraphics[width=8.0cm]{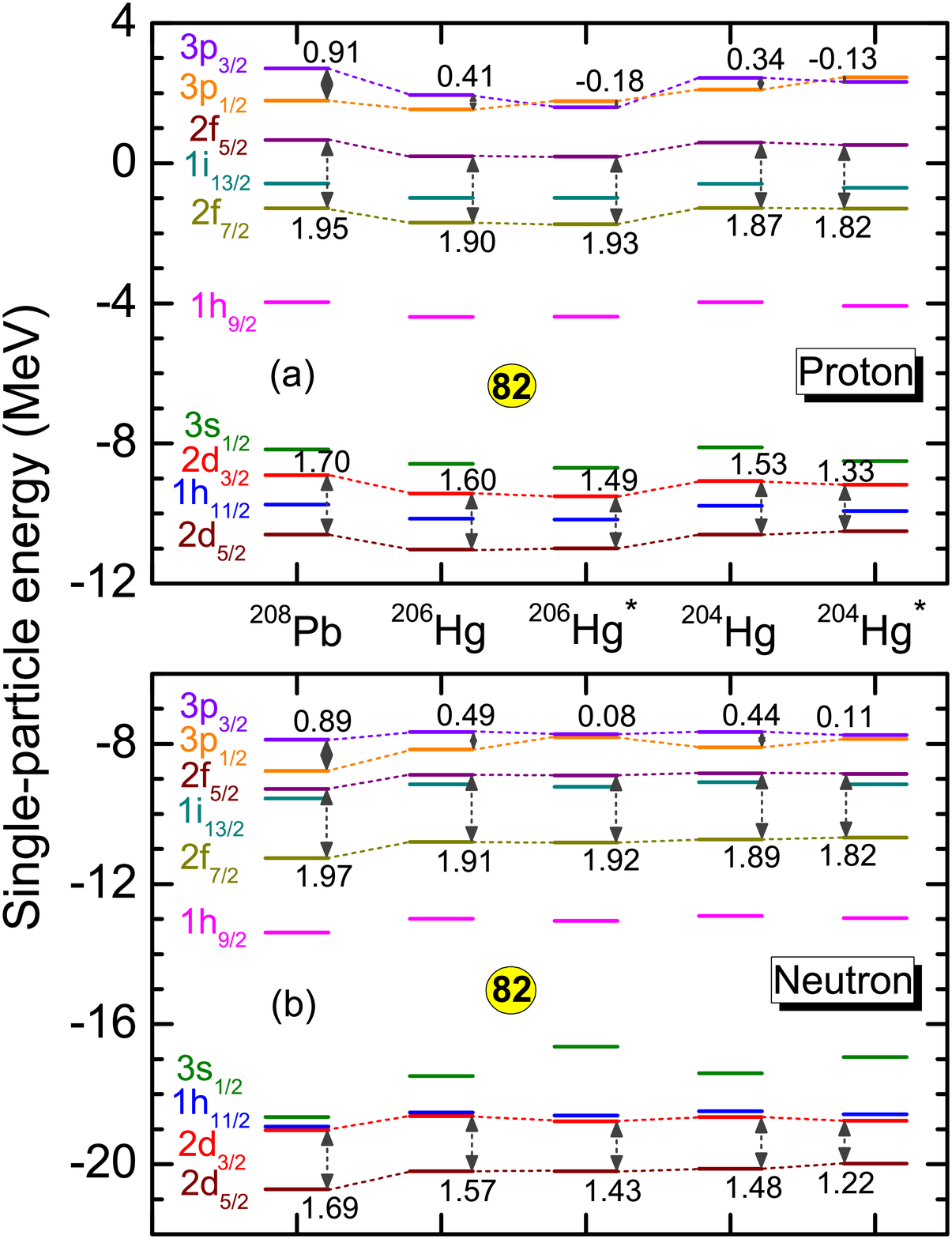}
 \caption{(Color online) (a) Proton and (b) neutron single-particle energies corresponding to the spherical configuration of $^{208}$Pb, $^{206}$Hg, and $^{204}$Hg from the mean-field calculations with the PC-PK1 force.
 The results without (w/o) pairing correlations, denoted with $^{206}$Hg$^\ast$ and $^{204}$Hg$^\ast$, are also given for comparison. The size of spin-orbit splitting is indicated with the value in units of MeV.}
 \label{SPE-HgPb}
 \end{figure}

 Figure~\ref{SPE-HgPb} displays the proton and neutron single-particle energies corresponding to the spherical states of $^{208}$Pb, $^{206}$Hg, and $^{204}$Hg by the PC-PK1 force. It is shown that the spin-orbit splitting is significantly quenched in the $3p$ partner states of $^{204, 206}$Hg, compared with those of $^{208}$Pb. Moreover, without performing the pairing correlations calculations, it can be seen that the discrete partners present almost two-fold degenerate not only in neutron but also in proton and even inversion of $\pi3p_{3/2}$ and $\pi3p_{1/2}$ orbitals.

 \begin{figure}[htbp]
 \centering
 \includegraphics[width=8.8cm]{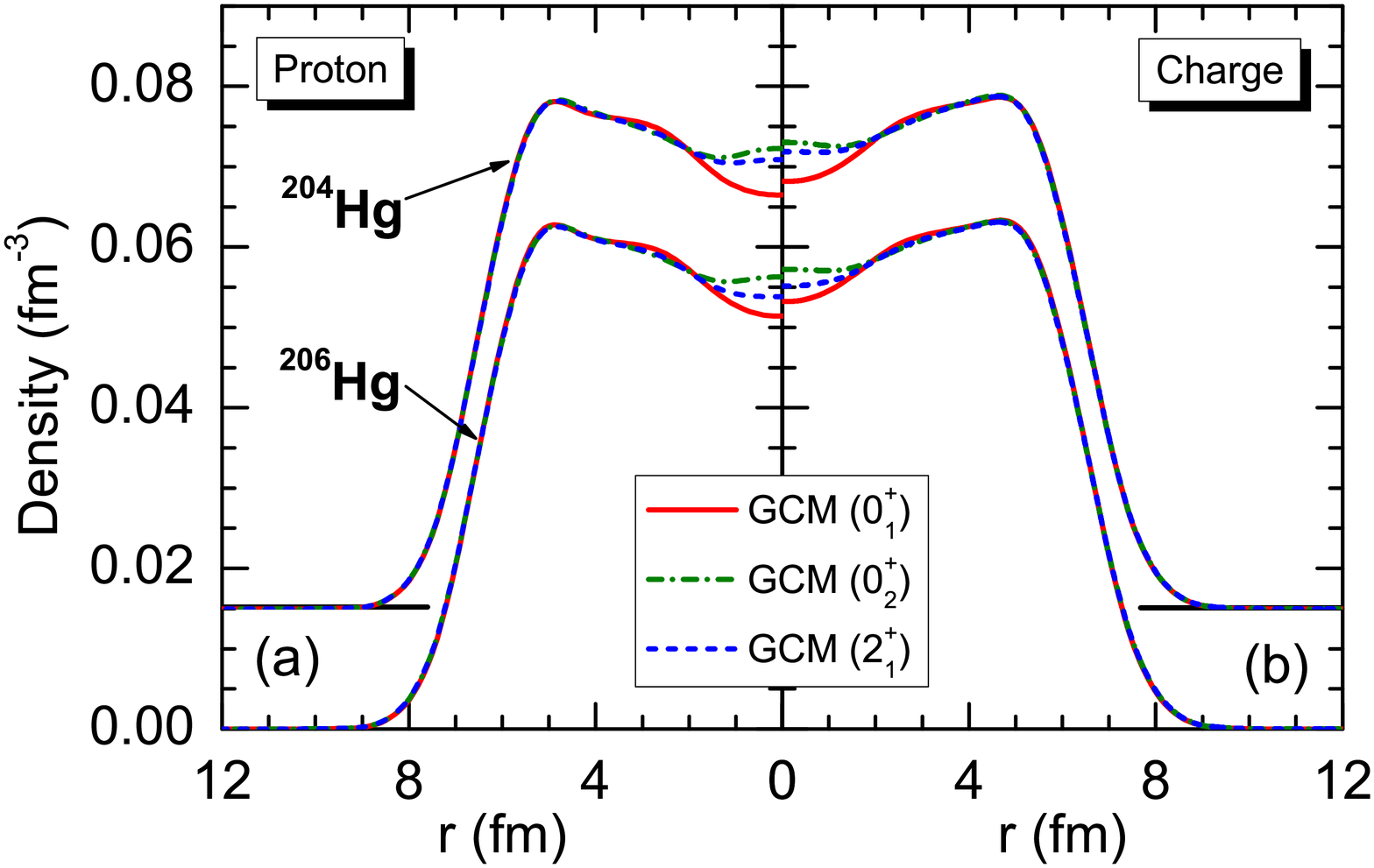}
 \caption{(Color online) Comparison between (a) proton and (b) charge density distributions for the $0^+_1, 0_2^+$, and $2_1^+$ states in $^{204, 206}$Hg. For $^{204}$Hg, the density distributions have been shifted up by 0.015 fm$^{-3}$.
 }
 \label{206Hggcmdens}
 \end{figure}

 \begin{table*}[t]
 \caption{
 \label{tab3}
 The central and maximal values of proton and charge densities (fm$^{-3}$), as well as the depletion factors $F_{max}^{\tau}$ and $F_{sat}^{\tau}$ [cf. Eq.~(\ref{Fmax})] of the $0_2^+$ and $2_1^+$ states in $^{204, 206}$Hg. }
 \begin{ruledtabular}
 \begin{tabular}{llllllllll}
  Nucleus       & State           &  $\rho_{cent}^p $ &  $ \rho_{cent}^{ch}$    &  $\rho_{max}^p$    &  $\rho_{max}^{ch}$     & $F_{max}^p$ &  $F_{max}^{ch}$  &  $F_{sat}^p$  &  $F_{sat}^{ch}$  \\
  \hline
   $^{204}$Hg  & GCM ($0_2^+$)    &  $0.057 $  &  $0.058 $       &  $0.063 $    &   $0.064 $         &  $0.10 $  &  $0.09 $     & $0.09 $   & $0.08 $        \\
               & GCM ($2_1^+$)    &  $0.056 $  &  $0.057 $       &  $0.063 $    &   $0.064 $         &  $0.12 $  &  $0.11 $     & $0.11 $   & $0.10 $        \\
  \hline
  $^{206}$Hg   & GCM ($0_2^+$)    &  $0.056 $  &  $0.057 $       &  $0.063 $    &   $0.063 $         &  $0.10 $  &  $0.10 $     & $0.09 $   & $0.08 $        \\
               & GCM ($2_1^+$)    &  $0.054 $  &  $0.055 $       &  $0.063 $    &   $0.063 $         &  $0.14 $  &  $0.13 $     & $0.13 $   & $0.11 $        \\

 \end{tabular}
 \end{ruledtabular}
 \end{table*}

 \subsection{Bubble structure in low-lying excited states}

 The previous studies~\cite{Yao13,Wu17a} demonstrate that the existence of bubble structure is unlikely in the low-lying excited states of  $^{34}$Si. It is interesting to discuss this noteworthy issue for $^{204, 206}$Hg. Figure~\ref{206Hggcmdens} displays the proton and charge density distributions of the $0_2^+$ and $2_1^+$ states in $^{204, 206}$Hg, in comparison with that of the ground state. One can see that the central depression in the excited states is less evident than that in the ground state. Even though the density around the center in the excited states  becomes flat, it is still much lower than the maximal density around 5.0 fm. Therefore, one still has a sizable value (around 0.1) for the  depletion factors $F_{max}^{\tau}$ and $F_{sat}^{\tau}$, cf. Table~\ref{tab3}. The decrease of the density from 5.0 fm towards 2.0 fm remains among the $0_1^+$, $0_2^+$, and $2_1^+$ states after taking into account the dynamic correlations, which is also exhibited in the ground state of
 $^{204, 206}$Hg. It means that this structure is formed mainly by the Coulomb repulsion, instead of the vacancy of the proton $3s_{1/2}$ orbit.


 \section{Summary}
 \label{Summary}
 We have reported a beyond mean-field calculations of the proton and charge distributions in the low-lying states of $^{204, 206}$Hg based on a relativistic point-coupling energy density functional. The dynamic correlations associated with symmetry restoration and shape mixing have been taken into account in the framework of particle-number and angular-momentum projected generator coordinate method. We have found that the dynamic correlations improve significantly the description of the charge-density difference between $^{208}$Pb and $^{204}$Hg.
 In contrast to the light nuclear systems, a semi-bubble structure is visible not only in the ground state of $^{204, 206}$Hg, but also in their $0_2^+$ and $2_1^+$ excited states. The bubble structure is rather robust under the perturbation of dynamic correlations. The results show that both the quantum shell effect and repulsive Coulomb interaction are responsible for the formation of the bubble structures in the nuclei of this mass region. In addition, it is worth mentioning that the dynamic correlations, tensor force or pairing correlations can modify the occupancy of the $s$ orbit around the Fermi surface and thus change the central bubble structure that is formed by the vacancy of the $s$ orbit. However, the bubble structure in heavy nuclei formed by the Coulomb repulsion effect can still survive with these effects.


 {\em Acknowledgments}. This work was supported in part by the National Natural Science Foundation of China under Grant Nos. 11275160, 11575148, 11475140, 11305134, 11765015, the Joint Fund Project of Education Department in Guizhou Province (No. Qian Jiao He KY Zi[2016]312), and by the Qiannan normal University Initial Research Foundation Grant to Doctor(qnsyrc201617).



\end{document}